\documentclass[aps,prb,twocolumn,english,superscriptaddress,citeautoscript,preprintnumbers,amsmath,amssymb,floatfix,footinbib]{revtex4-2}
\usepackage{float}
\usepackage[breaklinks=true,colorlinks,citecolor=blue,linkcolor=blue,urlcolor=blue]{hyperref}
\usepackage{amsmath}
\usepackage{graphicx}
\usepackage{dcolumn}
\usepackage{float}
\usepackage[utf8]{inputenc}
\usepackage{color}
\usepackage{placeins}

\graphicspath{{fig/}{./fig/}{.}}

\begin{document}
	
    \title{Electronic structure and anisotropic magnetotransport in the topological kagome ferromagnet MgMn$_6$Sn$_6$}
    \author{Debasmita Swain}
	\affiliation{Department of Physics, Indian Institute of Technology Kanpur, Kanpur 208016, India}
	\author{Mohit Mudgal}
	\affiliation{Research Institute for Synchrotron Radiation Science (HiSOR), Hiroshima University, Higashi-Hiroshima 739-0046, Japan}
    \author{Suman Nandi}
	\affiliation{Research Institute for Synchrotron Radiation Science (HiSOR), Hiroshima University, Higashi-Hiroshima 739-0046, Japan}
    \author{Aradhana Kumari}
	\affiliation{Research Institute for Synchrotron Radiation Science (HiSOR), Hiroshima University, Higashi-Hiroshima 739-0046, Japan}
    \author{Yogendra Kumar}
	\affiliation{Research Institute for Synchrotron Radiation Science (HiSOR), Hiroshima University, Higashi-Hiroshima 739-0046, Japan}
    \author{Vivek Kumar Malik}
	\affiliation{Department of Physics, Indian Institute of Technology Roorkee, Roorkee 247667, India}
    \author{Anup Pradhan Sakhya}
    \affiliation{Research Institute for Synchrotron Radiation Science (HiSOR), Hiroshima University, Higashi-Hiroshima 739-0046, Japan}
    \author{Masahiro Sawada}
    \affiliation{Research Institute for Synchrotron Radiation Science (HiSOR), Hiroshima University, Higashi-Hiroshima 739-0046, Japan}
    \author{Shin-ichiro Ideta}
    \affiliation{Research Institute for Synchrotron Radiation Science (HiSOR), Hiroshima University, Higashi-Hiroshima 739-0046, Japan}
    \author{Kenya Shimada}
    \affiliation{Research Institute for Synchrotron Radiation Science (HiSOR), Hiroshima University, Higashi-Hiroshima 739-0046, Japan}
    \affiliation{Research Institute for Semiconductor Engineering (RISE), Hiroshima University, Higashi-Hiroshima 739-8527, Japan}
    \affiliation{International Institute for Sustainability with Knotted Chiral Meta Matter (WPI-SKCM$^2$), Hiroshima University, Higashi-Hiroshima 739-8531, Japan}
	\author{Jayita Nayak}
    \thanks{Corresponding author:\href{mailto:jnayak@iitk.ac.in}{jnayak@iitk.ac.in}}\affiliation{Department of Physics, Indian Institute of Technology Kanpur, Kanpur 208016, India}

\begin{abstract}
We report the magnetic, magnetotransport, and electronic properties of the kagome ferromagnet MgMn$_6$Sn$_6$ using magnetization, angle dependent magnetoresistance, x-ray magnetic circular dichroism (XMCD), angle resolved photoemission spectroscopy (ARPES), and first-principles calculations. MgMn$_6$Sn$_6$ exhibits ferromagnetic ordering near $T_{\mathrm C}\approx295$ K with pronounced easy plane magnetic anisotropy. At low temperatures and low magnetic fields, the magnetoresistance (MR) is strongly anisotropic with respect to the magnetic field orientation, evolving from a predominantly negative MR for in-plane fields to a more quadratic behavior for out-of-plane fields. Angle dependent measurements further reveal a pronounced twofold MR anisotropy with additional higher order contributions. The finite orbital-to-spin moment ratio revealed by XMCD suggests a significant role of spin–orbit coupling (SOC) in MgMn$_6$Sn$_6$. The first-principles calculations show a Dirac-like band crossing at the $K$ point and a van Hove singularity (VHS) at the $M$ point, as expected for kagome materials. ARPES measurements resolve a sixfold symmetric Fermi surface and its systematic evolution with binding energy, in overall agreement with first-principles calculations. The measured band dispersions are also broadly consistent with the calculated multiband electronic structure. These results establish the connection between magnetic anisotropy, anisotropic magnetotransport, and the kagome derived electronic structure of MgMn$_6$Sn$_6$.
\end{abstract}
\maketitle
\section{INTRODUCTION}

The discovery of topological electronic states in kagome lattices has stimulated intense interest in condensed matter physics owing to the unique coexistence of geometric frustration, magnetism, and nontrivial band topology~\cite{Yin2022,Wang2023,Negi_2025}. The corner sharing network of transition metal atoms naturally hosts Dirac crossings, flat bands, and van Hove singularities~\cite{PhysRevB.109.035124,article,PhysRevB.87.115135}, which, in the presence of SOC and magnetic order, give rise to Berry curvature driven phenomena such as the anomalous Hall effect (AHE)~\cite{Liu2018,PhysRevLett.119.056601,Li2023}, topological Hall effect~\cite{PhysRevB.103.014416,PhysRevB.101.094404}, quantum oscillations with nontrivial Berry phase~\cite{PhysRevB.109.035124}, and Chern insulating states~\cite{,Yin2022,Yin_Ma_2020}. Consequently, magnetic kagome metals have emerged as an important platform for investigating the interplay between electronic topology and magnetism.\\
\indent Among the various kagome systems, the $R$Mn$_6$Sn$_6$ (R = rare earth, alkaline earth, or alkaline metal element) family has attracted particular attention as the Mn kagome bilayers exhibit strong exchange interactions~\cite{PhysRevX.12.021043,PhysRevB.103.014416,PhysRevB.103.235109}, while the choice of the $R$ element provides an effective means to tune magnetic anisotropy and electronic topology~\cite{PhysRevB.108.045132,PhysRevB.105.155108}. Depending on the $R$ ion, these compounds display a rich variety of magnetic ground states ranging from collinear ferromagnetism to ferrimagnetism and non-collinear antiferromagnetism~\cite{MAZET199954,MALAMAN199741}, accompanied by large anomalous Hall responses~\cite{PhysRevB.103.235109} and topological electronic bands. The strong coupling between the magnetic configuration and electronic structure makes this family an excellent platform for exploring topology driven magnetotransport~\cite{PhysRevB.108.045132,PhysRevB.105.155108}.\\
\indent MgMn$_6$Sn$_6$ is a particularly interesting member of this family as the nonmagnetic Mg layer eliminates the complexity associated with rare-earth magnetic moments~\cite{SONG2024172182,SONG2024101493}, enabling direct investigation of the intrinsic properties of the Mn kagome network. The compound orders ferromagnetically well above room temperature~\cite{Ma_2026,SONG2024172182} and exhibits strong easy plane magnetic anisotropy arising from SOC~\cite{Pal2025}. First-principles calculations have revealed multiple kagome derived bands crossing the Fermi level, suggesting a complex multiband electronic structure with potential topological character~\cite{h9ym-4cp2}. Recent ferromagnetic resonance (FMR) measurements have further demonstrated an unusually large magnetocrystalline anisotropy~\cite{Pal2025}, emphasizing the important role of SOC in determining both its magnetic and electronic properties. This raises an important question of how the pronounced magnetic anisotropy is reflected in the electronic structure and field dependent transport of MgMn$_6$Sn$_6$. Angle dependent magnetotransport measurements provide a sensitive means of probing such anisotropic responses and offer information complementary to conventional magnetic resonance techniques~\cite{PhysRevB.74.205205}.\\
\begin{figure*}[htbp!]
\centering
\includegraphics[width=\linewidth]{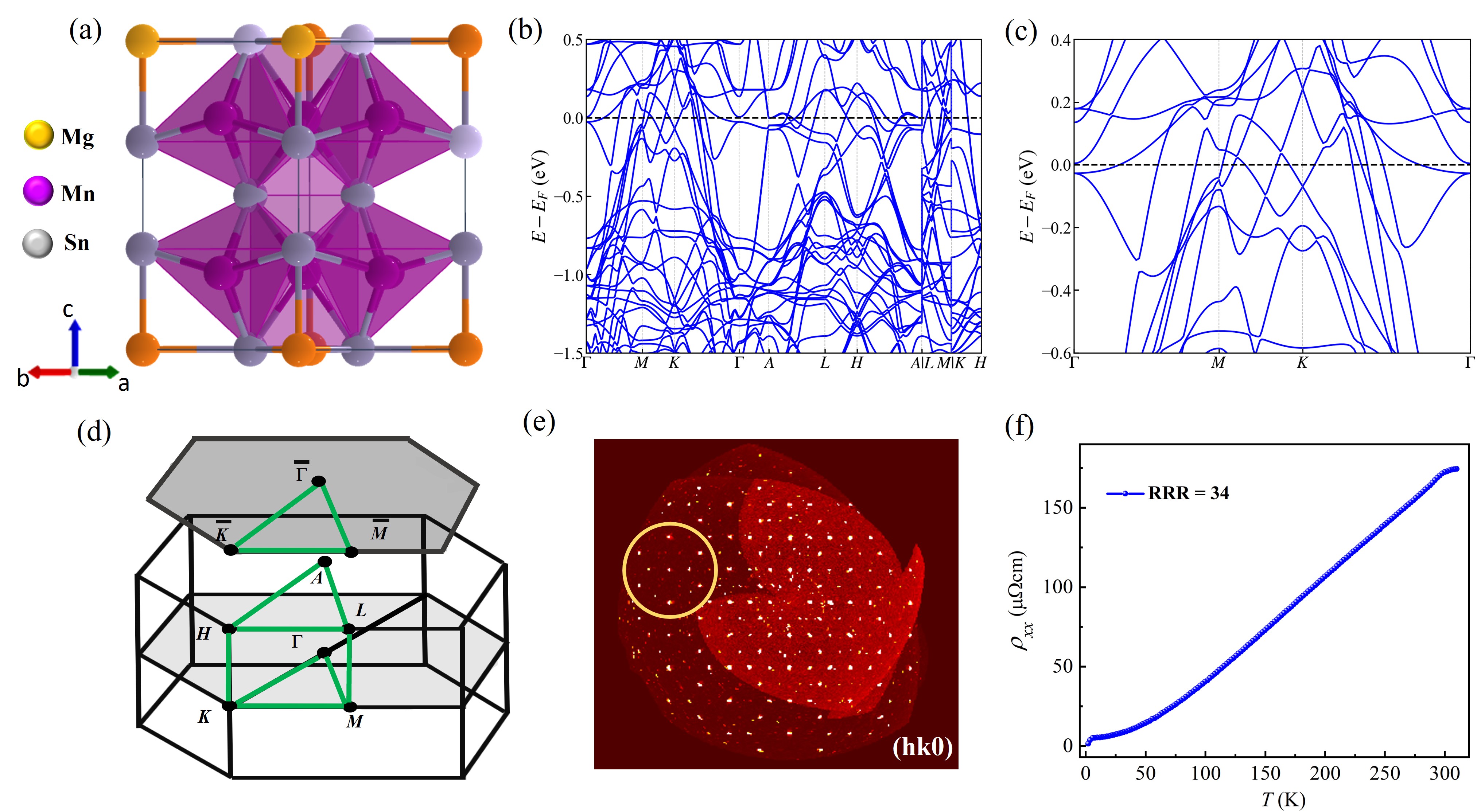}
\caption{Crystal structure and electronic structure of MgMn$_6$Sn$_6$. (a) Unit cell showing the primitive hexagonal structure, with Mg (golden), Mn (Purple) and Sn (silver) atoms. The crystal structure is constructed from the SC-XRD structure refined file using VESTA. (b) Calculated electronic band structure along high symmetry directions of the Brillouin zone, with the Fermi level set to zero energy (dashed line). (c) The zoomed-in calculated electronic band structure along the $\Gamma-M-K-\Gamma$ path. (d) Primitive bulk Brillouin zone with high symmetry points labeled. (e) SC-XRD precession image in the ($hk0$) plane showing 6-fold symmetry (as marked by the yellow circle). (f) Electrical resistivity as a function of temperature within a $T$ range of 2 K-310 K.}
\label{F1_Structural}
\end{figure*}
\indent Despite this growing interest on MgMn$_6$Sn$_6$, a direct, experimentally resolved picture of its electronic structure is still missing. This mirrors a broader pattern across the RMn$_6$Sn$_6$ (166) family as a whole, although dozens of these compounds have by now been synthesized and characterized by magnetometry, resistivity, and Hall effect measurements, only a handful (mainly YMn$_6$Sn$_6$~\cite{article} and TbMn$_6$Sn$_6$~\cite{Yin_Ma_2020} among them) have been examined by ARPES to directly resolve their kagome derived Dirac crossings, flat bands, and VHSs. 
 As a result, most of what is known about the electronic structure of this family, including MgMn$_6$Sn$_6$, still rests on first-principles calculations rather than direct spectroscopic measurement, leaving open whether the predicted band topology, and in particular its Fermi surface symmetry, is realized in the actual material. Addressing this gap for MgMn$_6$Sn$_6$ is one of the central motivations of the present work.\\
\indent In the present work, we report a comprehensive investigation of the  anisotropic magnetotransport  and magnetic properties of MgMn$_6$Sn$_6$ single crystals. Detailed magnetization measurements along both in-plane and out-of-plane directions reveal pronounced magnetic anisotropy. Magnetoresistance is investigated over a wide temperature range together with full $0^\circ-360^\circ$ angular dependent measurements and high resolution angular scans between $0^\circ$ and $90^\circ$. The finite orbital-to-spin moment ratio obtained from XMCD further reveals an appreciable orbital contribution to the Mn magnetic moment, suggesting a significant role of SOC in the magnetic and electronic properties of MgMn$_6$Sn$_6$.
First-principles calculations are employed to determine the electronic band structure and Fermi surface, and the calculated electronic topology broadly follows the measured ARPES results, demonstrating the characteristic sixfold symmetry of the kagome electronic states.
Our results establish a direct correlation between crystal symmetry, magnetic anisotropy, and anisotropic charge transport in MgMn$_6$Sn$_6$, providing new insight into the transport properties of kagome ferromagnets.
\section{Methods}\label{sec4}
MgMn$_6$Sn$_6$ single crystals were grown using a self-flux technique, as per the available reports~\cite{Ma_2026}. single crystal x-ray diffraction (SC-XRD) data were collected on a flat single crystal (Fig.\ref{F1_Structural}(e)) using a Bruker D8 Quest diffractometer equipped with a Photon II detector with Mo-K$\alpha$ radiation ($\lambda$ = 0.71073 $\AA$, graphite monochromator) at room temperature. The structure solution and refinement were carried out using Jana 2006 software~\cite{Palatinus:db5026,JANA_2006} by F$^2$ methods. The chemical composition was confirmed by energy dispersive x-ray spectroscopy (EDS) attached to a JEOL field emission scanning electron microscope. Magnetic measurements were carried out using a Quantum Design Magnetic Property Measurement System (MPMS, 6 T). Electrical transport measurements were performed in a Quantum Design Physical Property Measurement System (PPMS, 12 T) employing the conventional four probe technique with platinum wires attached using silver paste. To investigate anisotropic magnetotransport, angle dependent magnetoresistance measurements were carried out by rotating the sample with respect to the applied magnetic field over the full angular range of $0^\circ- 360^\circ$ at a fixed magnetic field of 12 T.\\
\indent The soft XAS and XMCD measurements were performed at the BL-14 beamline of the Hiroshima Synchrotron Radiation Center (HiSOR), Hiroshima University, Japan. Circularly polarized soft x-rays with a typical degree of circular polarization of approximately 0.7 were used to probe the Mn absorption edges. The measurements were carried out in the out-of-plane geometry, with the x-ray propagation direction and applied magnetic field parallel to the crystallographic $c$ axis, using the total electron yield (TEY) detection mode. XAS and XMCD spectra were acquired at 253 and 300 K under applied magnetic fields of 0.3 and 1.1 T. Due to the limitation of the beamline, measurements at fields above 1.1 T, as well as field-angular rotation studies, could not be performed. During the measurements, the sample was maintained under ultrahigh-vacuum conditions with a chamber pressure of approximately 10$^{-8}$ Torr.\\
\indent ARPES measurements were performed at the BL-1 beamline of the Research Institute for Synchrotron Radiation Science (HiSOR), Hiroshima University. The measurements were carried out using $p$-polarized synchrotron radiation with photon energies ranging from 120 to 150 eV. Single crystals of MgMn$_6$Sn$_6$ were cleaved \textit{in situ} under ultrahigh vacuum conditions better than $3 \times 10^{-9}$ Pa, exposing a clean surface for measurements. The ARPES spectra were acquired at 15 K temperature using a hemispherical electron analyzer (MBS-A1 analyzer). The energy and angular resolutions were set to approximately 20~meV and 0.1$^\circ$, respectively.\par
\textit{First principle calculations:}
First-principles calculations were performed based on density functional theory (DFT) using the Vienna \textit{ab initio} Simulation Package (VASP)~\cite{PhysRevB.54.11169,PhysRevB.59.1758}. The exchange correlation interaction was described within the generalized gradient approximation (GGA) using the Perdew-Burke-Ernzerhof functional~\cite{PhysRevLett.77.3865}. The magnetic moments of Mn were treated in a ferromagnetic configuration oriented along the $y$ direction, while SOC was incorporated self-consistently. A plane wave cutoff energy of 380 eV was used for all calculations. BZ sampling was carried out using an $11 \times 11 \times 7$ $k$-point mesh within the Gamma scheme. Based on the converged DFT results, maximally localized Wannier functions (MLWFs) were generated using the Wannier90 package~\cite{MOSTOFI2008685}. The corresponding Wannier-based tight binding Hamiltonian was subsequently used for additional calculations and for direct comparison with the ARPES measurements using WannierTools~\cite{WU2018405}.

\begin{figure*}[htpb!]
\centering
\includegraphics[width=0.9\linewidth]{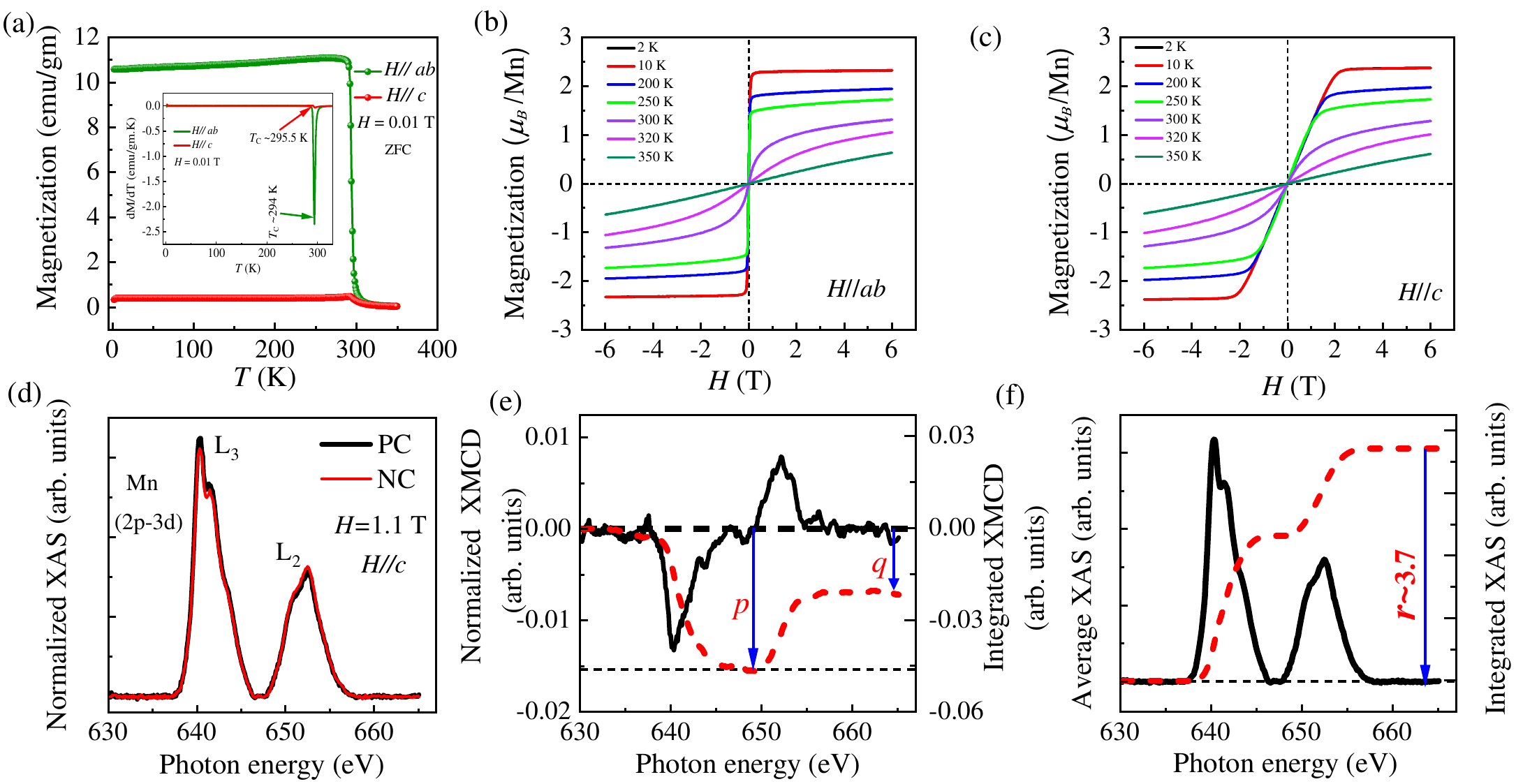}
\caption{Magnetic and XMCD properties of MgMn$_6$Sn$_6$. (a) Zero-field-cooled magnetization M(T) at $H$ = 0.01 T for $H \parallel ab$ and $H \parallel c$; inset shows dM/dT, from which $T_c$~$\approx$ 295 K is extracted. (b,c) Isothermal magnetization M($H$) at selected temperatures between 2 K and 350 K for $H \parallel ab$ and $H \parallel c$, respectively. (d) Normalized x-
ray absorption spectra (XAS) of the positive circular (PC) and negative circular (NC) polarization configurations at $H=1.1$~T and $T$=253 K. (e) Corresponding normalized XMCD spectrum at $253$~K together with its integrated curve, where $p$ and $q$ denote the integrated contributions used in the XMCD sum-rule analysis. (f) Sum of the XAS spectra and its integrated intensity, where $r$ is obtained from the total integrated XAS for the XMCD sum-rule analysis.}
\label{F2_Magnetization}
\end{figure*}

\section{CRYSTAL STRUCTURE AND DFT CALCULATIONS}
MgMn$_6$Sn$_6$ crystallizes in the hexagonal HfFe$_6$Sn$_6$-type structure (space group $P$6/${mmm}$), with Mn atoms forming kagome layers separated along $c$ axis by alternating Mg and Sn spacer layers as shown in Fig.~\ref{F1_Structural}(a). The calculated electronic band structure is shown in Fig.~\ref{F1_Structural}(b) and the zoomed-in calculated band structure along the $\Gamma-M-K-\Gamma$ path is given in Fig~\ref{F1_Structural}(c). The band structures show a Dirac-like band crossing at the $K$ point and a VHS at the $M$ point (see Fig.~\ref{F1_Structural}(b)), as expected for kagome materials. The bulk and (001) surface projected Brillouin zones (BZs) are shown in Fig.~\ref{F1_Structural}(d), with the relevant high symmetry points labeled to define the momentum space directions used for the band structure calculations and ARPES measurements discussed below. The (hk0) reciprocal space map from SC-XRD, as shown in Fig.~\ref{F1_Structural}(e), confirms the hexagonal symmetry together with the absence of twinning or secondary phases in the crystals selected for ARPES. The corresponding SC-XRD structural refinement, phase purity, and compositional analysis using (EDS) are presented in the Supplementary Material (SM, sec:S1).\\
\indent Temperature dependent resistivity, $\rho_{xx}(T)$ shown in Fig.~\ref{F1_Structural}(f), exhibits clean metallic behavior with a high residual resistivity ratio $\rho(300\,\mathrm{K})/\rho(5.5\,\mathrm{K}) \approx 34$, confirming the good quality of the single crystals used throughout this study. A weak downturn in the resistivity is observed below $\sim$ 3 K. Since the crystals were grown using Sn flux, this low temperature feature may arise from a minor contribution of residual Sn, whose superconducting transition occurs near 3.7 K. Similar low temperature resistive anomalies associated with residual Sn flux inclusions have been reported in other Sn flux grown intermetallics, including TbV$_6$Sn$_6$~\cite{PhysRevMaterials.6.104202} and La$_3$ZrSb$_5$~\cite{Khoury_2024}. A distinct anomaly is observed near $T_{\mathrm{C}} \approx 295$~K, corresponding to the ferromagnetic transition. Above $T_C$, the resistivity shows a weak tendency toward saturation that persists beyond 300~K, indicating only a marginal temperature dependence in the high temperature paramagnetic regime. The temperature dependent heat capacity data are provided in Fig.~S3 of the SM, where the lattice contribution approaches the Dulong–Petit limit of 3NR above $\sim220$ K.

\section{MAGNETIZATION MEASUREMENTS}
\indent Fig.~\ref{F2_Magnetization}(a) shows the zero field cooling (ZFC) magnetization curves for $H\parallel c$ and $H\parallel ab$ directions at an applied field of $H$ = 0.01 T. The inset shows the corresponding first order derivative plot, which clearly identifies a sharp ferromagnetic transition at $T_C$~$\approx$ 295 K, in good agreement with the Curie temperature reported from earlier studies~\cite{Ma_2026}. In addition, a pronounced easy plane anisotropy is observed, with the $H\parallel ab$ response being an order of magnitude larger than $H \parallel c$ throughout the ordered state. This confirms that the Mn moment lies predominantly within the kagome $ab$-plane.
This anisotropy is corroborated by the isothermal $M(H)$ data. For $H \parallel ab$, the low temperature (2–10 K) loops saturate rapidly reaching $\approx$ 2.2 $\mu_B$/Mn, and progressively lose both saturation and remanence on warming toward $T_C$ as shown in Fig.~\ref{F2_Magnetization}(b). Fig.~\ref{F2_Magnetization} (c) shows the magnetization for $H \parallel c$, where magnetization at the same temperatures increases much more gradually and remains unsaturated toward a comparable high field moment, indicating a sizable magnetocrystalline anisotropy field that must be overcome to rotate the moments out of the easy plane. To further investigate the magnetic response at the Mn site, element specific XMCD measurements were performed at the Mn L$_{2,3}$ edges.

\section{X-RAY MAGNETIC CIRCULAR DICHROISM (XMCD)}
\indent Figure~\ref{F2_Magnetization}(d) presents the Mn L$_{2,3}$ edge normalized XAS spectra of  of MgMn$_6$Sn$_6$ measured at 253~K in an out-of-plane ($H\parallel c$) geometry under an applied field of $H=1.1$~T. The XAS spectra were normalized to the average of L$_3$ edge peak intensity of the PC and NC polarization channels. A suitable background was subsequently subtracted by linearly interpolating between selected points in the spectra. The obtained XAS spectra exhibit distinct absorption features at the Mn  L$_3$ and L$_2$ edges at photon energy of $\approx640$ and $\approx652$~eV, respectively, with an intensity ratio of roughly $0.32:0.16$. L$_3$ edge shows a pronounced main peak accompanied by two weak higher energy shoulder features, while the L$_2$ edge displays a comparatively broader structure with indications of peak splitting. These features reflect the characteristic Mn 3$d$ electronic states and their multiplet related structure as observed in other Mn based systems~\cite{Sajedi2023,Watson_2013}. The corresponding XMCD spectrum is obtained by taking the difference of background subtracted NC--PC as shown in Fig.~\ref{F2_Magnetization}(e). The spectrum exhibits opposite sign features across the two edges, with a negative L$_3$ lobe reaching a minimum near $-0.016$ and a positive L$_2$ lobe peaking near $+0.010$. The energy-integrated form (red dashed curve, right axis) provides the sum-rule inputs $p$ and $q$, obtained from the value of the integral at the L$_3$/L$_2$ crossover ($\approx-0.046$) and at the high-energy plateau ($\approx -0.02$), respectively. The isotropic sum spectrum, as given in Fig.~\ref{F2_Magnetization}(f), and corresponding running integral provide the normalization constant $r$ (plateau $\approx3.7$). Together, $p$, $q$, and $r$ constitute the inputs for the Thole-Carra XMCD sum rules~\cite{PhysRevLett.68.1943},
\begin{equation}
m_{\mathrm{orb}}=-\frac{4}{3}\frac{q}{r}n_h\mu_B,
\end{equation}
and
\begin{equation}
m_{\mathrm{spin}}^{\mathrm{eff}}
=-2\frac{3p-2q}{r}n_h\mu_BC
=m_{\mathrm{spin}}+7m_T,
\end{equation}

\begin{figure}
\centering
\includegraphics[width=\linewidth]{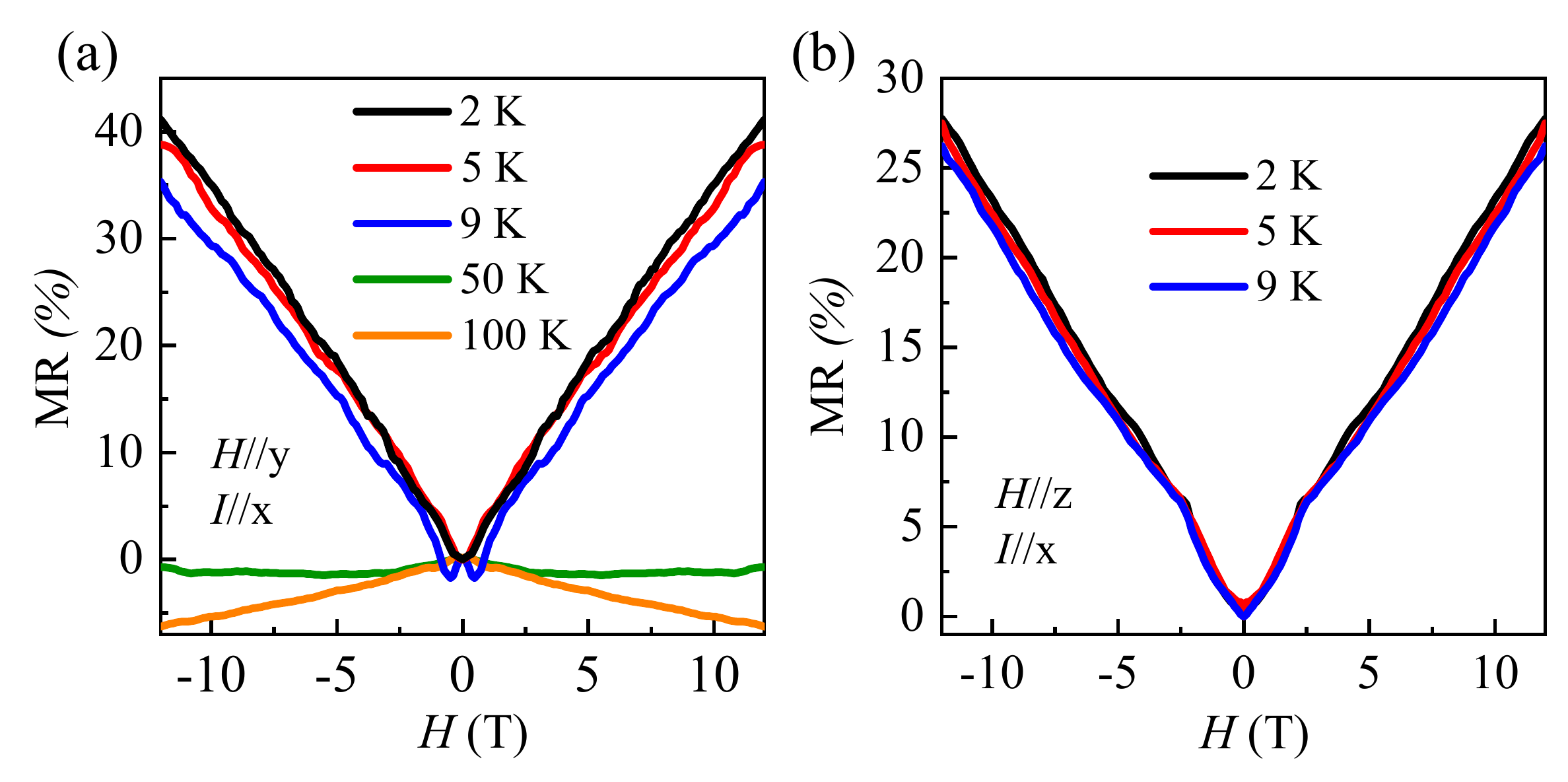}
\caption{Field-dependent MR, $\mathrm{MR}(H)=\left[\rho(H)-\rho(0)\right]/\rho(0)\times100\%$, measured at selected temperatures for (a) $H\parallel y$ and (b) $H\parallel z$.}
\label{MR}
\end{figure}

where $n_h$ is the number of $3d$ holes and here we assumed it to be five and $m_T$ is the magnetic-dipole term. To account for the significant j-j mixing between the Mn L$_{3}$ and L$_{2}$ edges, a correction factor of $C$ ($\approx$ 1.4) was incorporated into the spin sum rule analysis~\cite{PhysRevB.71.064418,Goering}. The effective spin and orbital magnetic moments per Mn atom are 0.374 $\mu_B$/Mn and 0.036 $\mu_B$/Mn, respectively, yielding an orbital-to-spin moment ratio of 9.6 \%. Such a sizable contribution of orbital moment suggesting incomplete quenching of Mn orbital moment and suggest a significant role of SOC in MgMn$_6$Sn$_6$ consistent with the strong magnetocrystalline anisotropy and its proposed SOC origin reported from FMR measurements~\cite{Pal2025}. Similar XMCD based analysis has previously been used in Fe$_3$Sn$_2$, where the enhanced orbital contribution relative to elemental Fe was interpreted as a quantitative manifestation of strong SOC and was further used to calibrate the SOC strength in DFT calculations~\cite{Zhang2024}.
The total obtained effective moment ($\approx$0.41 $\mu_B$/Mn) for the studied MgMn$_6$Sn$_6$ is slightly less than the value obtained from the isothermal magnetization data and this discrepancy arises as soft XMCD (and XAS) measurements in total electron yield mode are sensitive only to the sample's near surface region, whereas DC magnetization reflects the response of the bulk. Lowering the field to 0.3 T at the same temperature further reduces the net moment to 0.32 $\mu_B$/Mn (see SM, Fig.S4). When the sample is warmed just past the Curie temperature ($\approx$ 295 K) to 300 K, at a field of 1.1 T, the obtained moment is 0.26 $\mu_B$/Mn (see SM, Fig.S4), slightly less than the value measured at 253 K. This is consistent with the behaviour of typical ferromagnets near $T_C$, where the system remains in an intermediate state between ferromagnetic and paramagnetic phases slightly above $T_C$. Therefore an applied field can still induce short range, ferromagnetic-like spin correlations even above the transition~\cite{Watson_2013}. 
\begin{figure*}[t!]
\centering
\includegraphics[width=0.7\linewidth]{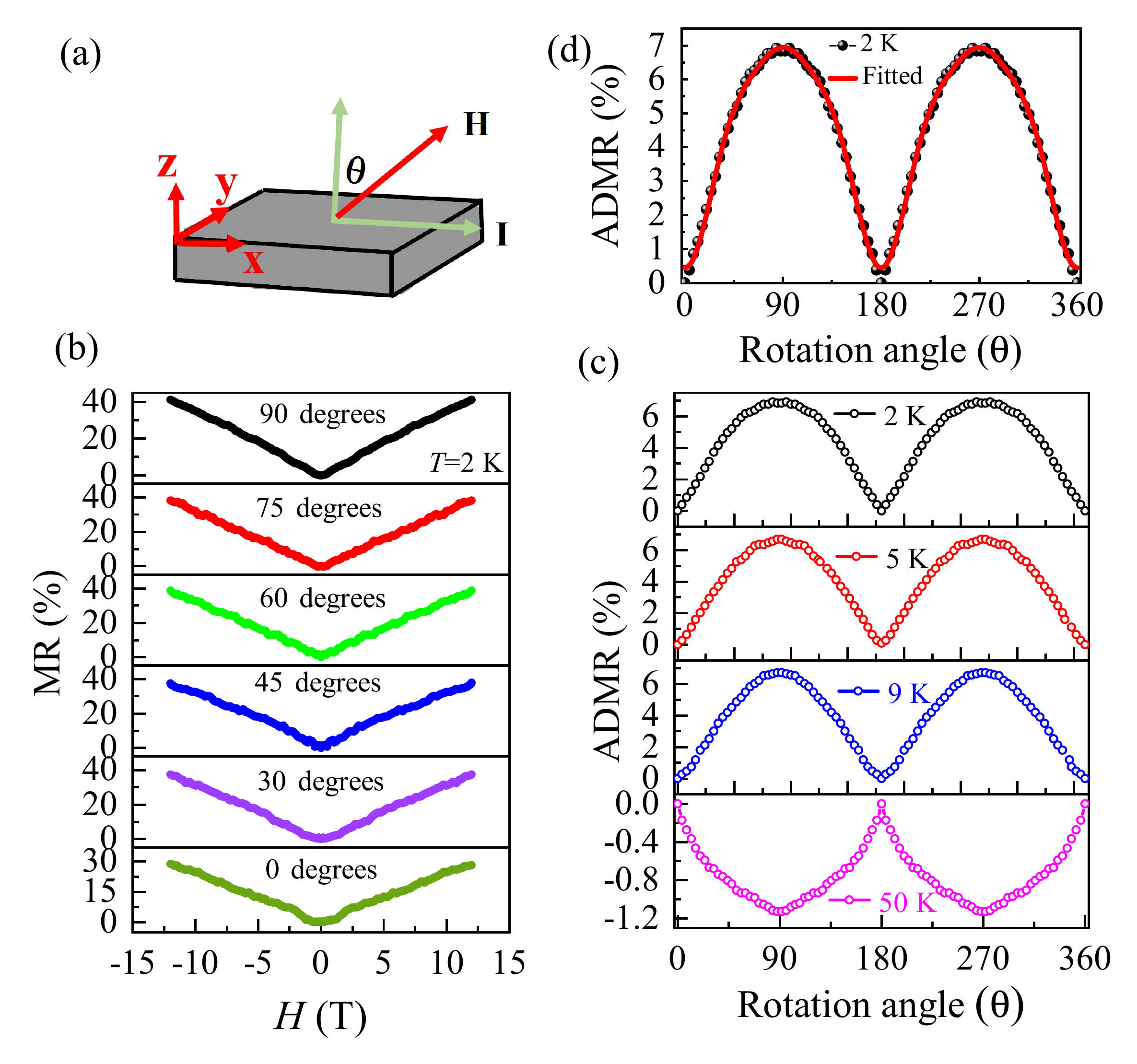}
\caption{(a) Schematic representation of the measurement configuration used for the angle-dependent studies. (b) Field dependent magnetoresistance (MR) measured at $T=2$~K for different rotation angles $\theta=0^\circ$, $30^\circ$, $45^\circ$, $60^\circ$, $75^\circ$, and $90^\circ$. (c) Angular dependence of the anisotropic magnetoresistance (ADMR) at $T=2$, $5$, $9$, and $50$~K under a magnetic field of 12~T. (d) ADMR at $T=2$~K (black solid balls) with the fitted curve given by the red solid line through the data is as per the equation mentioned in the corresponding text.}
\label{rotation}
\end{figure*}

The relatively weak dichroism signal compared to the isotropic absorption is consistent with the measurement geometry, which probes the magnetic hard axis ($c$ axis) rather than the easy $ab$-plane, at a field below the low temperature out-of-plane saturation field of MgMn$_6$Sn$_6$ (see Fig.~\ref{F2_Magnetization}(c)). However, the obtained magnetic moment in the transverse XMCD measurement agrees well with the magnetization data, motivating a further in detail study of the magnetic anisotropy in this system with an in-plane probe.

\section{MAGNETOTRANSPORT MEASUREMENTS}
\indent Figure~\ref{MR} shows the MR, $\mathrm{MR}(H)=\left[\rho(H)-\rho(0)\right]/\rho(0)\times100\%$, obtained after symmetrizing the raw $\rho(H)$ data, according to $\rho_{\mathrm{sym}}(H)=\left[\rho(H)+\rho(-H)\right]/2$, to eliminate any antisymmetric (Hall) contribution arising from slight field or contact misalignment. The symmetrized MR is plotted at selected temperatures for $H\parallel y$ and  $H\parallel z$. For $H \parallel y$, as shown in Fig.~\ref{MR}(a), the MR at 9~K exhibits a clear crossover from a low field negative response to a large positive, non-saturating MR at higher fields, whereas at 2~K, the MR initially approaches the zero-MR axis before increasing rapidly into the positive regime up to 41\% at 12 T.
The positive high field MR becomes dominant with increasing magnetic field, whereas the low field negative MR arises from the suppression of spin-disorder scattering. As the temperature is increased to 50 and 100 K, close to $T_C$, the orbital contribution is significantly weakened while spin disorder effects remain dominant, resulting in a negative MR over the entire measured field range. For $H \parallel z$, the MR remains positive over the entire measured field range and exhibits only a weak temperature dependence below 10 K as shown Fig.~\ref{MR}(b). At low fields, the MR follows an approximately quadratic field dependence up to~2.5 T, consistent with the saturation field seen in M($H$). This suggests that below saturation, magnetotransport is dominated by spin disorder and domain scattering from the canted Mn sublattice. As the field exceeds $H_{sat}$ (~2.5 T) and the sample becomes a single ferromagnetic domain, this scattering channel is suppressed and the transport instead reflects the intrinsic band structure. This gets reflected in the high field MR behaviour observed for this configuration, where the orbital contribution dominates throughout the higher field range, yielding a smooth, positive, non-saturating MR that reaches approximately $27$--$28\%$ at 12 T. This pronounced anisotropy is consistent with the strong easy plane magnetic anisotropy of MgMn$_6$Sn$_6$, for which a field applied along the hard axis modifies the magnetic configuration only gradually, thereby allowing the orbital contribution to remain the dominant component of the magnetotransport.

\begin{figure*}[htbp!]
\centering
\includegraphics[width=0.95\linewidth]{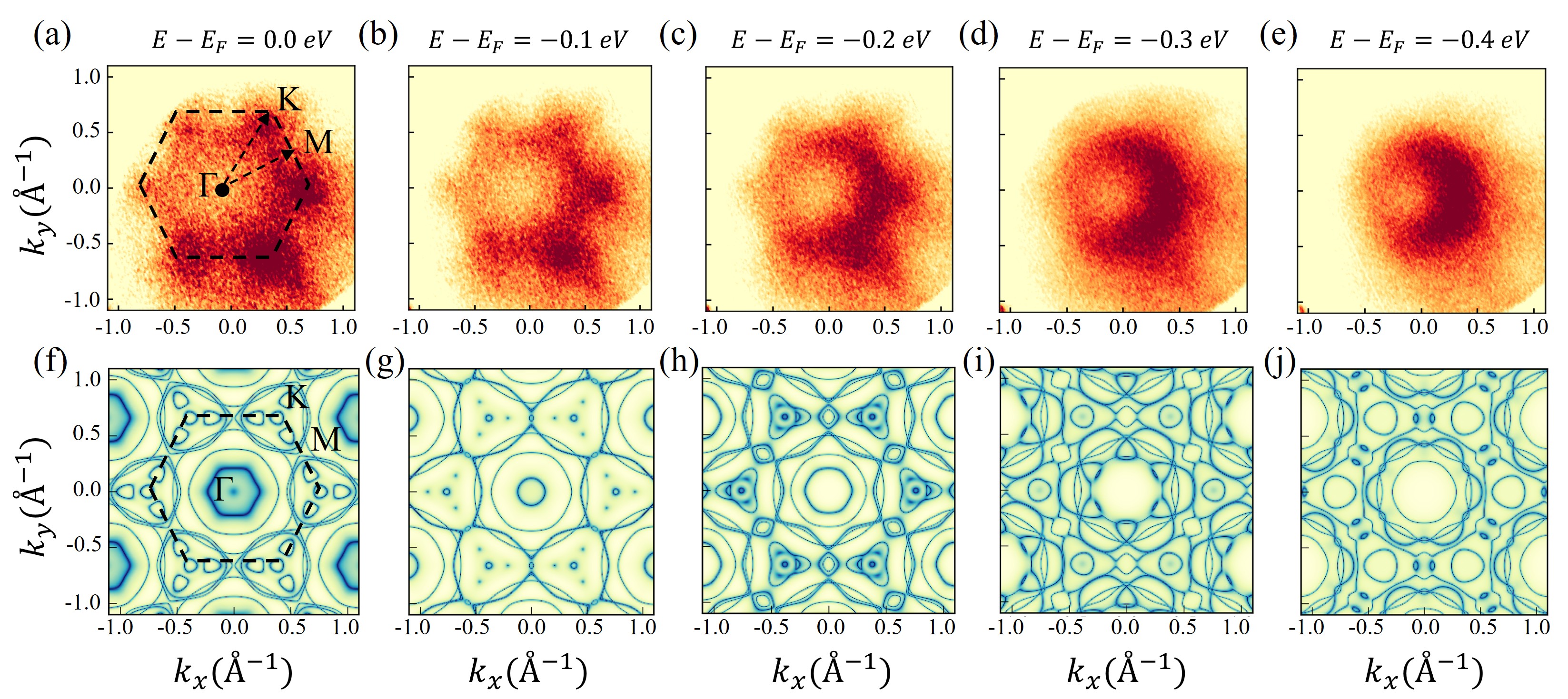}
\caption{Experimental and calculated constant energy contours of MgMn$_6$Sn$_6$. (a) ARPES constant energy maps measured at $T=15$ K using a photon energy of $h\nu=150$ eV at binding energies $E-E_F = 0.0, -0.1, -0.2, -0.3, -0.4$ eV.  (b) Corresponding calculated constant energy contours at the same binding energies, showing the evolution of the bulk electronic structure.
}
\label{FS}
\end{figure*}

The observed non-saturating linear MR in MgMn$_6$Sn$_6$ is unlikely to originate from disorder or inhomogeneity induced mechanisms as the measurements were performed on high quality single crystals, consistent with the interpretation proposed for HoMn$_6$Ge$_6$~\cite{PhysRevB.109.195104}. Instead, the DFT calculations reveal symmetry protected linear band crossings (Dirac/Weyl features) close to the Fermi level, suggesting that the linear MR is closely associated with the linearly dispersing electronic states near the Fermi energy with smaller effective mass~\cite{Abrikosov_2000}. This interpretation is further supported by reported theoretical studies showing that the Mn kagome lattice intrinsically hosts Dirac-like bands, largely independent of the rare earth ion~\cite{PhysRevB.108.045132}.

Figure~\ref{rotation}(b) shows the field dependence of the MR measured at $T$ = 2 K  over a magnetic field range of -12 T to +12 T, for field orientations $\theta = 0^\circ-90 ^\circ$, where $\theta$ denotes the rotation angle defined by the measurement geometric shown in Fig.~\ref{rotation}(a). In this measurement geometry, the crystal was rotated about the y-axis, continuously varying the field direction from the in-plane to the out-of-plane orientation. A nearly linear, non-saturating MR is observed for all field orientations up to $\theta = 45^\circ$. The MR reaches approximately $41\%$ at $\theta = 90^\circ$ and gradually decreases as the magnetic field is rotated from the in-plane easy-axis configuration toward the out-of-plane direction. At $\theta = 30^\circ$, the low field MR begins to deviate from the linear behavior and develops a noticeable quadratic field dependence, consistent with the behavior discussed in Fig.~\ref{MR}(b). This quadratic contribution becomes more pronounced at $\theta = 0^\circ$, where the maximum MR is reduced to approximately $28\%$. The crossover from predominantly linear to quadratic MR reflects the strong magnetic anisotropy of the system, highlighting the distinct responses of the easy and hard axis field configurations. This angular trend is pinned down more quantitatively in the ADMR sweeps of Fig.~\ref{rotation}(c).\\
\indent Figures~\ref{rotation}(c) displays the angular evolution of the ADMR at different temperatures and at fixed field of 12 T. The ADMR was measured by continuously rotating the sample through $360^\circ$. At low temperatures ($T \leq 9$~K), the ADMR exhibits a pronounced twofold symmetry with maxima located near $\theta \approx 90^\circ$ and $270^\circ$, and a broad minimum centered around $\theta$ $\approx$ $0^\circ(=360^\circ)$ and $180^\circ$. The ADMR reaches a maximum value of approximately $7\%$ at 2~K and gradually decreases with increasing temperature. The strong angular dependence of the MR reflects the underlying magnetic anisotropy of the system and the sensitivity of charge transport to the orientation of the magnetic moments relative to the applied field. Notably, the minimum in ADMR occurs when the field is aligned close to the hard axis direction, whereas enhanced MR is observed when the field is rotated toward the easy axis configuration.
\begin{figure*}[htbp!]
\centering
\includegraphics[width=0.7\linewidth]{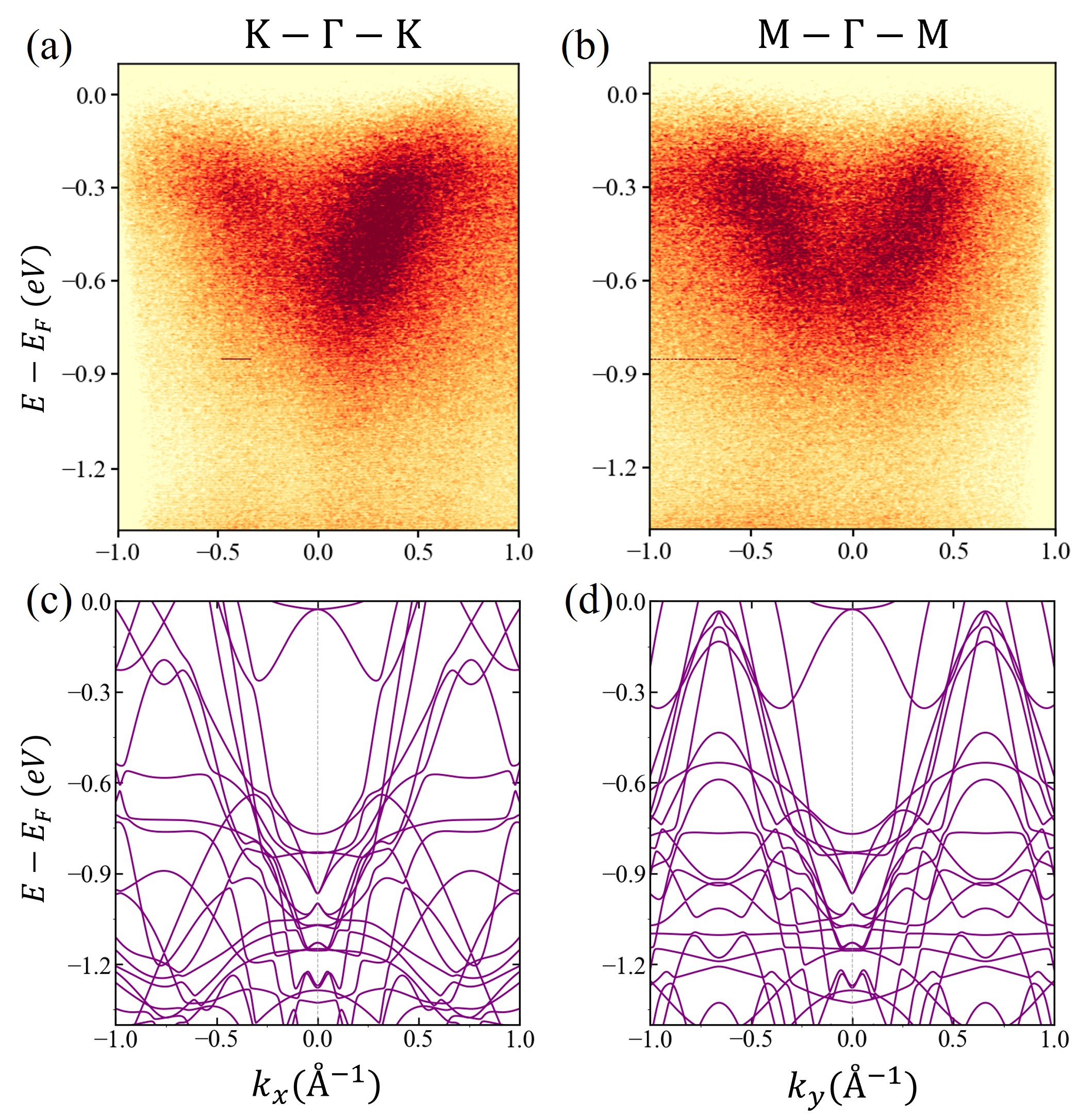}
\caption{Experimental and calculated electronic band structure of MgMn$_6$Sn$_6$. (a,b) ARPES intensity plots measured along the $K$–$\Gamma$–$K$ and $M$–$\Gamma$–$M$ high symmetry directions, respectively, at $T=15$ K using a photon energy of $h\nu=150$ eV. (c,d) Corresponding calculated electronic band structures along the same momentum directions.
}
\label{cuts}
\end{figure*}

To quantify the angular dependence, the low temperature ADMR data can be described by a combination of cosine harmonic terms of the form:
\[
\begin{aligned}
\mathrm{ADMR}(\theta) =\;& A_0 + A_2\cos(2\theta) \\
& + A_4\cos(4\theta) + A_6\cos(6\theta)
\end{aligned}
\]\\
Here, $A_0$ simply shows the temperature dependence of the zero field resistivity and $A_2$, $A_4$, and $A_6$ represent the two, four, and six-fold
symmetric coefficients, respectively.
The 2 K ADMR data fitted using the above equation is given in Fig.~\ref{rotation}(d), while the corresponding plots at $T$ = 5 K, 9 K, and 50 K are provided in Fig.S5 of SM. At 50~K, the angular profile deviates from the low-temperature behavior exhibiting a 90$^\circ$ phase shift from the cosine-like behavior. This evolution coincides with the crossover in the magnetoresistance, particularly the emergence of a negative MR for \(H\parallel y\) due to reduced spin scattering along the easy axis, suggesting a change in the relative contributions of the competing magnetic-scattering and orbital transport channels. The fitting of the ADMR data to the equation indicates that the magnetotransport is predominantly governed by twofold anisotropy with additional higher-order contributions upto 6$^{th}$ harmonics. The small sixfold component observed in the ADMR is likely associated with the underlying magnetocrystalline anisotropy. However, a detailed in-plane field rotation study would be helpful to establish its precise contribution. The twofold symmetry of the ADMR data is the usual term that originates from the relative angle between the applied in-plane current and the external field. The four-fold symmetric component observed in the ADMR is not anticipated on the basis of the point-group symmetry of the hexagonal lattice, which permits only two-fold and six-fold angular contributions. One possible origin is an angular modulation of the density of states near the Fermi level~\cite{GHOSH2025183506}. Such anomalous four-fold contributions, despite the underlying hexagonal symmetry, have also been reported in other 166-family kagome compounds~\cite{GHOSH2025183506} as well as in several other systems~\cite{Song2024Mn3Ga,Xiang2021,Li2023,PhysRevResearch.2.022029}. Within the scope of the present study, the microscopic origin of this term cannot be conclusively established, and is therefore left for a more detailed investigation in future work. Temperature evolution of the ADMR FFT amplitudes peaks at 60$^\circ$ (with C$_6$ symmetry), 90$^\circ$ (C$_4$ symmetry), and 180$^\circ$ (C$_2$ symmetry) are provided in the SM, Fig.S6. The good agreement between the experimental data and the fitted curves confirms that the angular response is dominated by the intrinsic anisotropic magnetotransport of the system.  

\section{FERMI SURFACE AND CONSTANT ENERGY CONTOURS}
To directly investigate the electronic structure of MgMn$_6$Sn$_6$, we performed ARPES measurements at $T=15$ K using a photon energy of $h\nu=150$ eV. Figs.~\ref{FS}(a-e) shows the measured constant energy contours at selected binding energies $E-E_F = 0.0, -0.1, -0.2, -0.3, -0.4$ eV, while the corresponding calculated contours are presented in Figs.~\ref{FS}(f-j). The experimental maps reveal a clear sixfold symmetric electronic structure centered around the $\bar{\Gamma}$ point, consistent with the underlying hexagonal symmetry of the Mn kagome lattice. With increasing binding energy, the momentum space contours undergo a pronounced evolution, reflecting the presence of multiple dispersive bands contributing over the measured energy range.\\
\indent A comparison with the calculated constant energy contours shows good overall agreement in both the symmetry and the evolution of the electronic states. In particular, the calculations reproduce the central $\bar{\Gamma}$-centered features together with the surrounding sixfold symmetric intensity distribution observed experimentally. The agreement between experiment and calculation confirms that the observed ARPES features predominantly originate from the intrinsic bulk electronic structure of MgMn$_6$Sn$_6$ and provides direct spectroscopic evidence for the characteristic hexagonal electronic topology expected from its kagome derived band structure.

\section{ELECTRONIC BAND STRUCTURE ALONG HIGH SYMMETRY DIRECTIONS}

Figures \ref{cuts}(a) and \ref{cuts}(b) show the ARPES intensity distributions of MgMn$_6$Sn$_6$ measured along the $K$–$\Gamma$–$K$ and $M$–$\Gamma$–$M$ high symmetry directions, respectively, at $T=15$ K using a photon energy of $h\nu=150$ eV. The corresponding calculated band structures are shown in Figs. \ref{cuts}(c) and \ref{cuts}(d). Additionally, at 120 eV photon energy, the experimental Fermi surface and high-symmetry cuts are shown in Fig.~S7 of the SM. The ARPES spectra exhibit a broad V-shaped band with its bottom at about 0.9 eV below $E_F$ at the $\Gamma$ point. In Fig.~\ref{cuts}(a), the V-shaped band disperses upward towards $K-\Gamma$ while in Fig.~\ref{cuts}(b), it gradually flattens at the $M$ point. However, owing to the broad spectral features observed in the ARPES measurements, the topological features predicted by theoretical calculations, such as the linear band crossing and Van Hove singularity, cannot be conclusively established within the present study. A detailed investigation of these features is therefore left for future studies.

\section{CONCLUSION}
\indent In conclusion, we have presented a combined magnetization, angle-dependent magnetotransport, XMCD, ARPES, and first-principles investigation of the kagome ferromagnet MgMn$_6$Sn$_6$. Magnetization measurements establish ferromagnetic order below $T_C\approx295$~K with pronounced easy-plane anisotropy, which is directly reflected in the anisotropic magnetotransport response. For $H\parallel y$, the magnetoresistance evolves from a negative low-field response associated with spin-disorder scattering to a large, non-saturating positive MR reaching $\sim 41\%$ at 12~T, whereas for $H\parallel z$ it remains positive and reaches $\sim 28\%$ at 12~T. The angle-dependent MR exhibits a dominant twofold anisotropy with higher-order harmonics at the available temperatures, highlighting the close interplay between magnetic anisotropy and charge transport. XMCD measurements at the Mn $L_{2,3}$ edges reveal a finite orbital moment, $m_{\mathrm{orb}}\approx0.036~\mu_B/\mathrm{Mn}$, relative to $m_{\mathrm{spin}}\approx0.374~\mu_B/\mathrm{Mn}$, demonstrating a non-negligible orbital contribution and an appreciable role of SOC in the Mn derived states. Complementary ARPES measurements resolve a sixfold symmetric Fermi surface centered at $\Gamma$, in overall agreement with the calculated electronic structure.
Taken together, these results reveal the relationship between the crystal symmetry, magnetic anisotropy, and anisotropic charge transport in MgMn$_6$Sn$_6$, while suggesting a possible role of SOC in shaping the underlying electronic structure. These findings provide experimental insight into the interplay between magnetism and electronic structure in this kagome ferromagnet. Although the relatively broad ARPES features currently prevent an unambiguous band by band assignment of the states near $E_F$, higher resolution ARPES measurements combined with complementary probes of the in-plane anisotropy will be important for resolving the individual band crossings and establishing the full extent of SOC-driven anisotropy in MgMn$_6$Sn$_6$.
\begin{acknowledgements}
D.S.  acknowledges the financial support of Anusandhan National Research Foundation (ANRF), India, under the National Post-Doctoral Fellowship (NPDF) scheme file No. PDF/2025/004781. The ARPES measurements and XMCD measurements were performed with the approval of the Proposal Assessing Committee of the Research Institute for Synchrotron Radiation Science (Proposal No. 26AU017 and Proposal No. 26AU016, respectively). We thank N-BARD, Hiroshima University, for supplying the liquid helium. The work at Hiroshima University was supported by JST EXPERT-J, Japan Grant Number JPMJEX2510. We acknowledge IIT Kanpur, Science and Engineering Research Board (project no.
CRG/2023/00786060 and SPG/2021/000443), and the Department of Science and Technology for financial support.
\end{acknowledgements}
\section*{AUTHOR CONTRIBUTIONS}
J.N. and D.S. conceived the idea of the manuscript. J.N supervised the project and provided overall leadership, coordination, and integration across different research activities. D.S. synthesizes, characterizes the single crystals, and performs magnetic and transport measurements. V.K.M. provided the magnetic measurement facilities. M.M. performed the ARPES measurements with assistance from Y.K., S.N., A.P.S., S.I.I., and K.S. A.K. and M.M. performed XMCD measurements with the help of M.S. M.M. analyzed the ARPES data. S.N. performed the DFT calculations. D.S. and M.M. prepared the manuscript with contributions from all authors. All authors discussed the results and contributed to the final version of the manuscript.

\vspace{0.5cm}
\textit{Note added.} During the preparation of this manuscript, another work was published on arXiv which reports anomalous magnetotransport in MgMn$_6$Sn$_6$ with non-collinear magnetic structure and anisotropic Hall response~\cite{deb2026anomalousmagnetotransportnoncollinearcorrelated}.

\section*{Data Availability Statement}
The datasets generated during and/or analyzed during the current study are available from the corresponding author on reasonable request.
 
\section*{Competing interests}
The authors declare no competing interests.


\bibliography{biblio.bib}
\end{document}